\documentstyle[epsfig]{mn2e}

\def\gtsima{$\; \buildrel > \over \sim \;$}
\def\ltsima{$\; \buildrel < \over \sim \;$}
\def\gtrsim{\lower.5ex\hbox{\gtsima}}
\def\lesssim{\lower.5ex\hbox{\ltsima}}

\begin{document}

\title[Dynamics of MSBHs in YSCs and ULXs]{Dynamics of massive stellar black holes in young star clusters and the displacement of ultra-luminous X-ray sources}
\author[Mapelli et al.]
{M. Mapelli$^{1}$, E. Ripamonti$^{1}$, L. Zampieri$^{2}$, M. Colpi$^{1}$
\\
$^1$Universit\`a Milano Bicocca, Dipartimento di Fisica G.Occhialini, Piazza
della Scienza 3, I--20126, Milano, Italy; {\tt mapelli@mib.infn.it}\\
$^2$INAF-Osservatorio astronomico di Padova, Vicolo dell'Osservatorio 5, I--35122, Padova, Italy\\
}
\maketitle \vspace {7cm }

  \begin{abstract}
In low-metallicity environments, massive stars might avoid supernova explosion and directly collapse, forming massive ($\sim{}25-80\,{}M_\odot{}$) stellar black holes (MSBHs), at the end of their life. MSBHs, when hosted in young massive clusters, are expected to form binaries and to strongly interact with stars, mainly via three-body encounters. We simulate various realizations of young star clusters hosting MSBHs in hard binaries with massive stars. We show that a large fraction ($\sim{}44$ per cent) of MSBH binaries are ejected on a short timescale ($\le{}10$ Myr). The offset of the ejected MSBHs with respect to the parent cluster is consistent with observations of X-ray binaries and ultra-luminous X-ray sources. Furthermore, three-body encounters change the properties of MSBH binaries: the semi-major axis changes by $\le{}50$ per cent and the eccentricity of the system generally increases. We shortly discuss the implications of our simulations on the formation of high-mass X-ray binaries hosting MSBHs.

\end{abstract}
\begin{keywords}
black hole physics -- stars: binaries: general -- galaxies: star clusters: general -- X-rays: binaries
\end{keywords}

%

\section{Introduction}
According to models of stellar evolution (Fryer 1999; Heger et al. 2002, 2003), a star with a final mass\footnote{We name `final mass', $m_{\rm fin}$, of a star the mass bound to the star immediately before the collapse.}  $m_{\rm fin}$  higher than $m_{\rm th}\sim{}40$ M$_\odot{}$ can avoid supernova explosion and directly collapse into a black hole (BH). In this case, it is reasonable to expect that the mass of the remnant is comparable to the final mass of the progenitor star (or at least more than half of it). Therefore, BHs may form, via this channel, with mass higher than in the standard scenario of supernova explosion and fallback. According to Belczynski et al. (2010), the masses of BHs formed via direct collapse may be as high as $\sim{}80$ M$_\odot{}$. In the following, we will refer to massive stellar BHs (MSBHs) to indicate BHs with mass $25-80$ M$_\odot{}$ formed via direct collapse.

The final mass of a star likely depends on metallicity, as mass losses by stellar winds correlate with metallicity (see Bresolin \&  Kudritzki 2004 for a review). Massive stars with (about) solar metallicity cannot have a final mass as high as $\sim{}40$ M$_\odot{}$, because they lose a large fraction of their mass by stellar winds. Thus, solar-metallicity stars never form BHs with mass higher than $\sim{}10-20$ M$_\odot{}$. Instead, stars with sufficiently low metallicity might have final masses $\gtrsim{}40$  M$_\odot{}$ and directly collapse into MSBHs. The metallicity threshold for the formation of MSBHs depends on the adopted model of stellar wind and stellar evolution.  For example, assuming the model by Belczynski et al. (2010), based on the stellar-wind model by Vink, de Koter \&{} Lamers (2001), the threshold metallicity is Z$_{\rm th}\approx{}0.4$ Z$_\odot{}$. We stress that the dependence of stellar winds on metallicity is quite uncertain, especially for massive stars, as it relies upon observations of a few massive stars, and it is drawn from models generally neglecting binary evolution and rotation (e.g., Bresolin \&{} Kudritzki 2004, and references therein). For this reason, there is an uncertainty of $\gtrsim{}0.2$ dex on Z$_{\rm th}$.

Recent papers (Mapelli, Colpi \&{} Zampieri 2009; Zampieri \&{} Roberts 2009; Mapelli et al. 2010a, 2010b) indicate that MSBHs born from the direct collapse of massive low-metallicity stars might account for a  fraction of ultra-luminous X-ray sources (ULXs). ULXs are  non-nuclear point-like sources with (isotropic) X-ray luminosity $L_{\rm X}\gtrsim{}10^{39}$ erg s$^{-1}$  (see Mushotzky 2004 for a review, and references therein), corresponding to the Eddington luminosity of a $\gtrsim{}7$ M$_\odot{}$ BH. 
The mechanism that powers the ULXs is still unknown, although various scenarios have been proposed. 
Apart from the aforementioned interpretation (i.e., that a fraction of ULXs may be powered by MSBHs formed from the evolution of massive metal-poor stars), 
ULXs could be associated with high-mass X-ray binaries (HMXBs) powered by stellar-mass black holes (BHs) with anisotropic X-ray emission (e.g. King et al. 2001) or with super-Eddington accretion rate/luminosity (e.g. Begelman 2002; King \&{} Pounds 2003; Socrates \&{} Davis 2006; Poutanen et al. 2007), or emitting via a combination of the two mechanisms (e.g. King 2008). Recently, Linden et al. (2010, hereafter L10) suggested that a fraction of ULXs may be explained with  Roche-lobe overflow (RLO) HMXBs in moderately low-metallicity environments, undergoing mildly super-Eddington accretion. In their scenario, the bulk of HMXBs hosts 10$-$15 M$_\odot{}$ BHs, but up to $\sim{}20$ per cent of  HMXBs hosts BHs with mass $>20$ M$_\odot{}$.
In addition, ULXs could also be associated with HMXBs powered by intermediate-mass BHs (IMBHs), i.e. BHs with mass $100\,{}{\rm M}_\odot{}\le{}m_{\rm BH}\le{}10^5\,{}{\rm M}_\odot{}$ (see van der Marel 2004 for a review).  
 Among the aforementioned hypotheses, beamed emission is problematic for those ULXs surrounded by isotropically ionized nebulae (Pakull \&{} Mirioni 2002; Kaaret, Ward \&{} Zezas 2004a) and for the brightest ULXs. Similarly, super-Eddington emission requires uncommonly high violations of the Eddington limit for the brightest ULXs. 
 On the other hand, the very existence of IMBHs lacks observational evidences and IMBHs are not needed to explain the observational properties of most of the ULXs (e.g. Gon\c{c}alves \&{} Soria 2006; Stobbart, Roberts \&{} Wilms 2006; Copperwheat et al. 2007; Roberts 2007;  Zampieri \&{} Roberts 2009). Furthermore, ULXs might actually be a mixed bag of different objects, including sources of different nature.


In the following, we will focus on the scenario proposed by Mapelli et al. (2009) and by Zampieri \&{} Roberts (2009), according to which a fraction of ULXs might 
be powered by MSBHs, accreting in HMXBs. This idea is supported by the fact that most  ULXs are located in galaxies with a high star formation rate (SFR, e.g.  Irwin, Bregman \&{} Athey  2004) and are often associated with low-metallicity environments (Pakull \&{} Mirioni 2002; Zampieri et al. 2004; Soria et al. 2005;  Swartz, Soria \&{} Tennant 2008; but see Winter, Mushotzky \&{} Reynolds 2007 for a different result\footnote{Winter et al. (2007) derive the metallicity of 16 ULXs from {\it XMM Newton} spectra and find approximately solar abundances for all of them. However, X-ray photo-ionization edges are known to be highly uncertain metallicity indicators. The abundances derived by Winter et al. (2007) are much higher ($0.2-1$ dex) than those derived, for the same galaxies, from spectra of HII regions using the P-calibration method (see, e.g., Pilyugin, Vilchez \&{} Contini 2004). For a more detailed discussion about metallicity estimators, see, e.g., Kennicutt, Bresolin \&{} Garnett (2003).}). Considering a sample of 66 late-type galaxies, Mapelli et al. (2010a,b) showed that the number of observed ULXs per galaxy ($N_{\rm ULX}$) scales almost linearly with the SFR and that $N_{\rm ULX}$/SFR (i.e. the number of observed ULXs per galaxy, normalized to the SFR) anti-correlates with the metallicity. Finally, the expected number of MSBHs per galaxy ($N_{\rm BH}$), derived in  Mapelli et al. (2010a,b), correlates with $N_{\rm ULX}$.

Thus, the idea that MSBHs are the engine of most ULXs agrees, under many respects, with the data. However, many  issues need to be addressed, in order to check this scenario. First, observations indicate that ULXs are generally found close to star forming regions and to young star clusters (YSCs), but often displaced from them, by a distance $\approx{}0.1-1$ kpc (e.g. Zezas et al. 2002; Swartz, Tennant \&{} Soria 2009; Berghea 2009; Swartz 2010). The same seems to occur for bright ($\ge{}10^{36}$ erg s$^{-1}$) X-ray sources in starburst galaxies (e.g. Kaaret et al. 2004b). This fact has been generally interpreted as the indication that bright X-ray sources are powered by runaway binaries, i.e. by binaries which were ejected from the parent cluster because of a natal kick (e.g. Sepinsky, Kalogera \&{} Belczynski 2005; Zuo \&{} Li 2010) or a close encounter (e.g. Kaaret et al. 2004b; Berghea 2009). 
 The amount and the distribution of natal kicks are very uncertain for BHs (e.g., Brandt \&{} Podsiadlowski 1995; Brandt, Podsiadlowski \&{} Sigurdsson 1995; Gualandris et al. 2005; Fragos et al. 2009, and references therein).
 It is reasonable to assume that natal kicks in the absence of supernova are smaller than in the supernova scenario (e.g. Fryer \&{} Kalogera 2001; L10). If the natal kick is lower than the escape velocity from the parent cluster, it may not be sufficient to eject the MSBH from the parent cluster (e.g., L10). L10 even assume that no natal kick is present in the direct collapse scenario, because of the absence of supernova explosion.
While the role of natal kicks needs to be investigated in more detail, we focus here on the effects of recoil velocities induced by three-body encounters: are they able to eject MSBHs?

 A second issue is related to the evolutionary path for the formation of HMXBs with MSBHs.  L10 suggest that a kick is almost necessary for a binary to enter the RLO--HMXB regime. According to their model, MSBHs can hardly enter the RLO--HMXB regime, as they have no natal kicks.
However, the model by L10 does not include three-body encounters. In this paper, we want to study the effects of  three-body encounters on the orbital parameters of a MSBH binary and on the possibility that such binary enters the RLO--HMXB regime.

Therefore, the aim of this paper is to check whether there may be an alternative source of kick, different from the natal one, for MSBH binaries. In particular, we check, via direct N-Body simulations, the occurrence and the consequences of kicks by three-body interactions involving MSBHs in YSCs.

We do not consider only MSBHs, but, for comparison, also intermediate-mass BHs (IMBHs). The plan of the paper is the following. In Section~2, we describe the simulations. In Section~3, we discuss the evolution of the radial distribution of the simulated MSBHs. In Section~4, we mention the effects of the kick on the orbital parameters of the involved binaries. In Section~5, we present simulations of IMBHs embedded in YSCs, for comparison with MSBHs. In Section~6, we compare our simulations with the available data in the literature. In Section~7, we report our conclusions.




\section{Simulations}
The simulations were done using the Starlab\footnote{\tt http://www.sns.ias.edu/$\sim{}$starlab/} software environment (see Portegies Zwart et al. 2001), which allows to integrate the dynamical evolution of a star cluster, resolving binaries and three-body encounters.

\begin{figure}
\center{{
\epsfig{figure=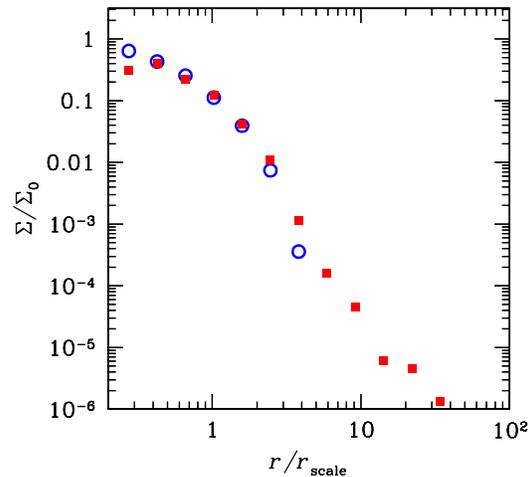,height=7cm} 
}}
\caption{\label{fig:fig1}
Surface density of the cluster ($\Sigma{}$), normalized to the central surface density value $\Sigma{}_0$, as a function of the two-dimensional projected distance $r$. $r_{\rm scale}$ is the scale radius and is $\sim{}1$ pc for all simulated clusters. Open circles (blue on the web): initial conditions; filled squares (red on the web): evolution after 10 Myr.}
\end{figure}

We make 25 realizations of isolated star clusters having a King profile with adimensional central potential W$_0$=5 (King 1966). Each cluster hosts N$_{\rm star}=5000$ stars, corresponding to a total mass $\approx{}3-4\times{}10^3$ M$_\odot{}$. We set  an initial binary fraction $f_{\rm bin}=0.1$. Stars are generated according to the Salpeter (1955) initial mass function (IMF), in a mass range from 0.08 M$_\odot{}$ up to 120 M$_\odot{}$. The virial radius of each cluster is $\approx{}$ 1 pc. Thus, the clusters have a half-mass relaxation time $t_{\rm rlx}\sim{}10$ Myr and a core collapse time scale $t_{\rm cc}\simeq{}0.2\,{}t_{\rm rlx}\approx{}1$ Myr (see Portegies Zwart \&{} McMillan 2002). We choose to simulate clusters with the same global properties to reduce the considered parameter space. No external tidal field is added. Stellar and binary evolution were switched off during these runs, to avoid introducing too many parameters in this first study. The effects of binary evolution will be taken into account in paper II.
Fig.~\ref{fig:fig1} shows the initial one-dimensional profile of one of the realizations, compared with the profile at $t=10$ Myr.
According to Mapelli et al. (2010a), clusters with these properties are expected to host $\approx{}1$ MSBH, if their metallicity is $\sim{}0.1$ Z$_\odot{}$. Therefore, we generate one MSBH in each simulated cluster. The mass of the MSBH is fixed to be 50 M$_\odot{}$, as fiducial value. 
The MSBH is always generated in a binary system, with a massive companion. The assumption that the MSBH is always in a binary is quite realistic, as the progenitor of the MSBH and the MSBH itself are among the most massive objects in the parent cluster (see, e.g., Kulkarni, Hut \&{} McMillan 1993).  The mass of the companion is uniformly distributed (following the standard settings of Starlab) between a minimum mass $m_{\rm low}$ and a maximum mass $m_{\rm high}$. We require that $m_{\rm low}=10$ M$_\odot{}$ and  $m_{\rm high}=50$ M$_\odot{}$. In fact, 
theoretical studies (e.g., Portegies Zwart, Dewi \&{} Maccarone 2004; Patruno et al. 2005; Madhusudhan et al. 2006) indicate that, for masses of the donor $m_2\lesssim{}10$ M$_\odot{}$, luminosities in the ULX range cannot be attained during main sequence (MS), except for very short outbursts ($\sim{}$ a few days every few months of quiescence). We do not consider ULXs with lower-mass evolved donors, because we are interested only in young systems ($t\lesssim{}10$ Myr), where stars less massive than $\sim 10 M_\odot$ did not yet  evolve off the MS and transfer mass steadily on the MSBH.
 Furthermore, we note that if the mass of the donor is larger than the mass of the BH, the system will have an unstable mass transfer (Rappaport, Podsiadlowski \&{} Pfahl 2005).  For this reason, we avoid companion masses larger than the MSBH mass.
The initial position and velocity of the centre of mass of the MSBH binary are drawn from the same distributions as the other stars in the cluster (see Portegies Zwart et al. 2001).

The initial semi-major axis and the initial eccentricity of the MSBH binary 
follow the same distributions adopted for `normal' binaries
 (i.e. a thermal distribution for eccentricities and a logarithmic distribution for semi-major axes, Portegies Zwart et al. 2001). We then exclude all MSBH systems with semi-major axis $a>10$ A.U. and $a<0.1$ A.U. and all systems that were already ionized or that exchanged companion after one timescale (corresponding to $\sim{}0.25$ Myr). We consider only systems with $a>0.1$ A.U., to exclude binaries whose semi-major axis is smaller than the tidal radius $r_{\rm tid}$ of the companion (for the binaries in our sample, $r_{\rm tid}=0.03-0.1$ A.U.). Furthermore, we select only binaries with $a<10$ A.U., because we are interested only in systems that are sufficiently hard to survive for most of the cluster's life\footnote{A binary is hard when $a\leq{}G\,{}m_1\,{}m_2\,{}(m_1+m_2+m_3)/\left[(m_1+m_2)\,{}m_3\,{}\sigma{}^2\right]$, where $G$ is the gravitational constant, $m_1$ and $m_2$ are the mass of the primary and of the secondary, respectively, $m_3$ is the average mass of a cluster star, $\sigma{}$ is the velocity dispersion of the cluster. For our systems, the MSBH binary is hard if $a\lesssim{}10$ A.U.}. Therefore, our sample is not representative of all possible binaries hosting MSBHs, but only of hard systems. 
We will consider softer binaries, whose dynamics is completely different, in a forthcoming paper.
 We also stress that our simulations start when the MSBH is already formed and do not make any assumptions about the dynamical and stellar evolution before the formation of the MSBH. Therefore, the simulated hard MSBH binaries are not necessarily primordial, but can have a dynamical origin (via exchanges).
Furthermore, our selection of the initial semi-major axes provides initial periods that are consistent with the expected distribution of periods of hard BH binaries in dense star clusters (e.g., figure~2 of Belczynski et al. 2004).
Each cluster realization has been integrated for 10 Myr, i.e. approximately the lifetime of a 15 M$_\odot{}$ star. This guarantees that the companion of the MSBH is still unevolved in most cases.

For comparison with the runs including MSBHs, we also simulated a set of 20 clusters with the same properties as before but hosting an IMBH (with mass 300 M$_\odot{}$) instead of a MSBH. The IMBH is always generated in a binary, with similar constraints to those adopted for MSBHs (i.e. companion mass between 10 and 120 M$_\odot{}$ and semi-major axis such that the binary is hard). 

\section{Close encounters and ejection}
\begin{figure}
\center{{
\epsfig{figure=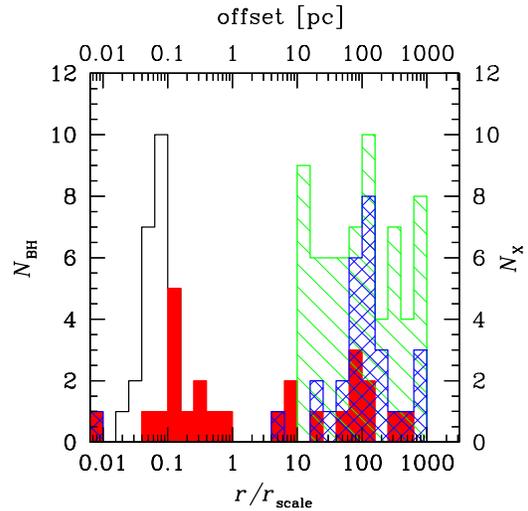,height=7cm} 

}}
\caption{\label{fig:fig2}
Empty black histogram: two-dimensional radial projected distribution of MSBHs at time $t=0$. Filled histogram (red on the web): two-dimensional  radial projected distribution of MSBHs at time $t=10$ Myr. The empty black histogram and the filled histogram refer to left-hand $y-$axis (i.e. the number of simulated MSBHs, $N_{\rm BH}$) and to the bottom $x-$axis (i.e. the two-dimensional distance $r$ of the MSBH from the centre of the parent cluster, normalized to the scale radius $r_{\rm scale}$. For all the simulated clusters  $r_{\rm scale}\sim{}1$ pc). Cross-hatched histogram (blue on the web): radial projected distribution of observed ULXs from the closest star-forming region (from the sample in table~9 of Berghea 2009). Hatched histogram (green on the web): radial projected distance of observed X-ray sources (with X-ray luminosity $L_X\ge{}5\times{}10^{35}$ erg s$^{-1}$) from the closest star cluster (considering M~82, NGC~1569, NGC~5253, from figure 3 of Kaaret et al. 2004b). Cross-hatched and hatched histograms  refer to right-hand $y-$axis (i.e. the number of observed X-ray sources, $N_{\rm X}$) and to the top $x-$axis (the offset, i.e. the projected distance of the X-ray source from the closest star cluster and/or star forming region, in pc). 
}
\end{figure}

Fig.~\ref{fig:fig2} shows the radial projected distribution of MSBHs at $t=0$ (empty black histogram) and at $t=10$ Myr (filled histogram, red on the web). The position of each MSBH in Fig.~\ref{fig:fig2} comes from a different run. We assume that a MSBH is definitely ejected from its parent cluster, when its projected distance from the centre of the cluster is larger than $4\,{}r_{\rm scale}$ (i.e. $\sim{}4$ times the virial radius, which corresponds approximately to the tidal radius of the clusters) and when its three-dimensional velocity is larger than 5 km s$^{-1}$ (i.e., the escape velocity from the parent cluster). According to this criterion, 11 MSBHs are ejected at $t\le{}10$ Myr, i.e. 44 per cent of the simulated MSBHs. Before the ejection, the MSBHs had a strong interaction with a single star, whose mass ranges from $\sim{}24$ M$_\odot{}$ to $\sim{}61$ M$_\odot{}$.
The ejected MSBHs are still in binaries (82 per cent, i.e. 9 runs) or even in triple systems (18 per cent, i.e. two runs), as it is reported in Table~1. This is a consequence of the fact that we consider only hard binaries. Furthermore, a non-negligible fraction of ejected MSBHs (27 per cent) exchanged their companion with a different star before the ejection.
Table~2 shows the analogous statistics for MSBHs that, at $t=10$ Myr, are still bound to the cluster. The percentage of ionizations among MSBHs that remain in the cluster is still very low, but the percentage of (dynamically unstable) triple systems is much larger than that of binaries. This means that most of MSBHs that remain inside the cluster are still interacting. Interestingly, no exchanges occur among the MSBHs that remain inside the cluster. 
\begin{figure}
\center{{
\epsfig{figure=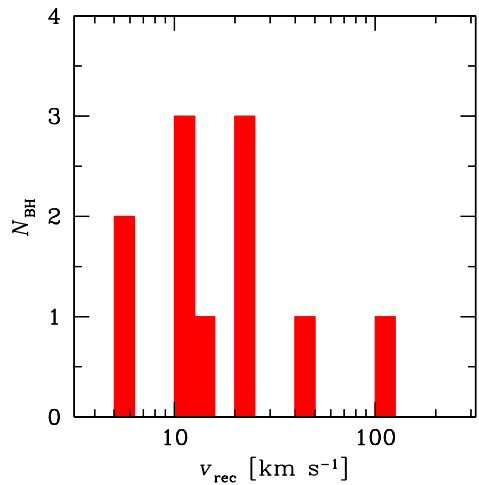,height=7cm} 
}}
\caption{\label{fig:fig3}
Distribution of three-dimensional recoil velocities ($v_{\rm rec}$) of the ejected MSBHs.
}
\end{figure}
\begin{figure}
\center{{
\epsfig{figure=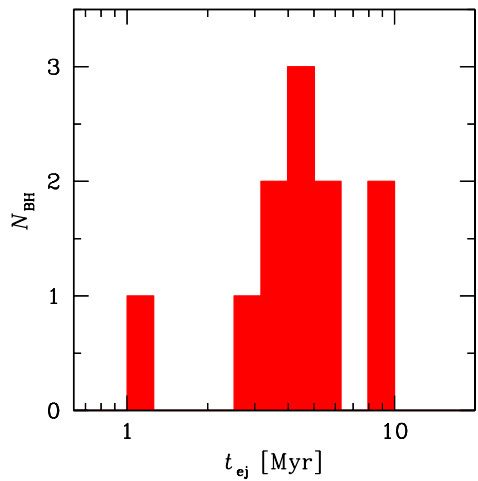,height=7cm} 
}}
\caption{\label{fig:fig4}
Distribution of ejection times of MSBHs from the cluster ($t_{\rm ej}$).
}
\end{figure}

Let us now focus on the properties of ejected MSBH systems.
Fig.~\ref{fig:fig3} shows the distribution of recoil velocities ($v_{\rm rec}$) of the ejected MSBHs. $v_{\rm rec}$ spans from 5 km s$^{-1}$ up to 120  km s$^{-1}$: it is, on average, smaller than the natal kick velocities assumed for stellar-mass BHs in L10 (see e.g. their fig.~6), but it is a viable substitute to natal kick for BHs born from direct collapse. Furthermore, velocities larger than $\sim{}100$ km s$^{-1}$ are likely to disrupt binaries (e.g., Berghea 2009).

\begin{table}
\begin{center}
\caption{End states of the simulated MSBHs that are ejected before $t=10$ Myr.} \leavevmode
\begin{tabular}[!h]{ll}
\hline
State 
& Percentage$^{\rm a}$\\
\hline
Ionization & 0.0\\
Binary & 81.8\\
Triple & 18.2\\
\noalign{\vspace{0.1cm}}
Exchange & 27.3\\
\noalign{\vspace{0.1cm}}
\hline
\end{tabular}
\footnotesize{\\$^{\rm a}$The percentage refers to the total of ejected MSBHs before $t=10$ Myr (i.e. 11 cases).}
\end{center}
\end{table}

\begin{table}
\begin{center}
\caption{End states of the simulated MSBHs that, at $t=10$ Myr, are still bound to the cluster.} \leavevmode
\begin{tabular}[!h]{ll}
\hline
State 
& Percentage$^{\rm a}$\\
\hline
Ionization & 7.2\\
Binary & 21.4\\
Triple & 71.4\\
\noalign{\vspace{0.1cm}}
Exchange & 0.0\\
\noalign{\vspace{0.1cm}}
\hline
\end{tabular}
\footnotesize{\\$^{\rm a}$The percentage refers to the total of MSBHs that are still inside the parent cluster at $t=10$ Myr (i.e. 14 cases).}
\end{center}
\end{table}

\begin{figure*}
\center{{
\epsfig{figure=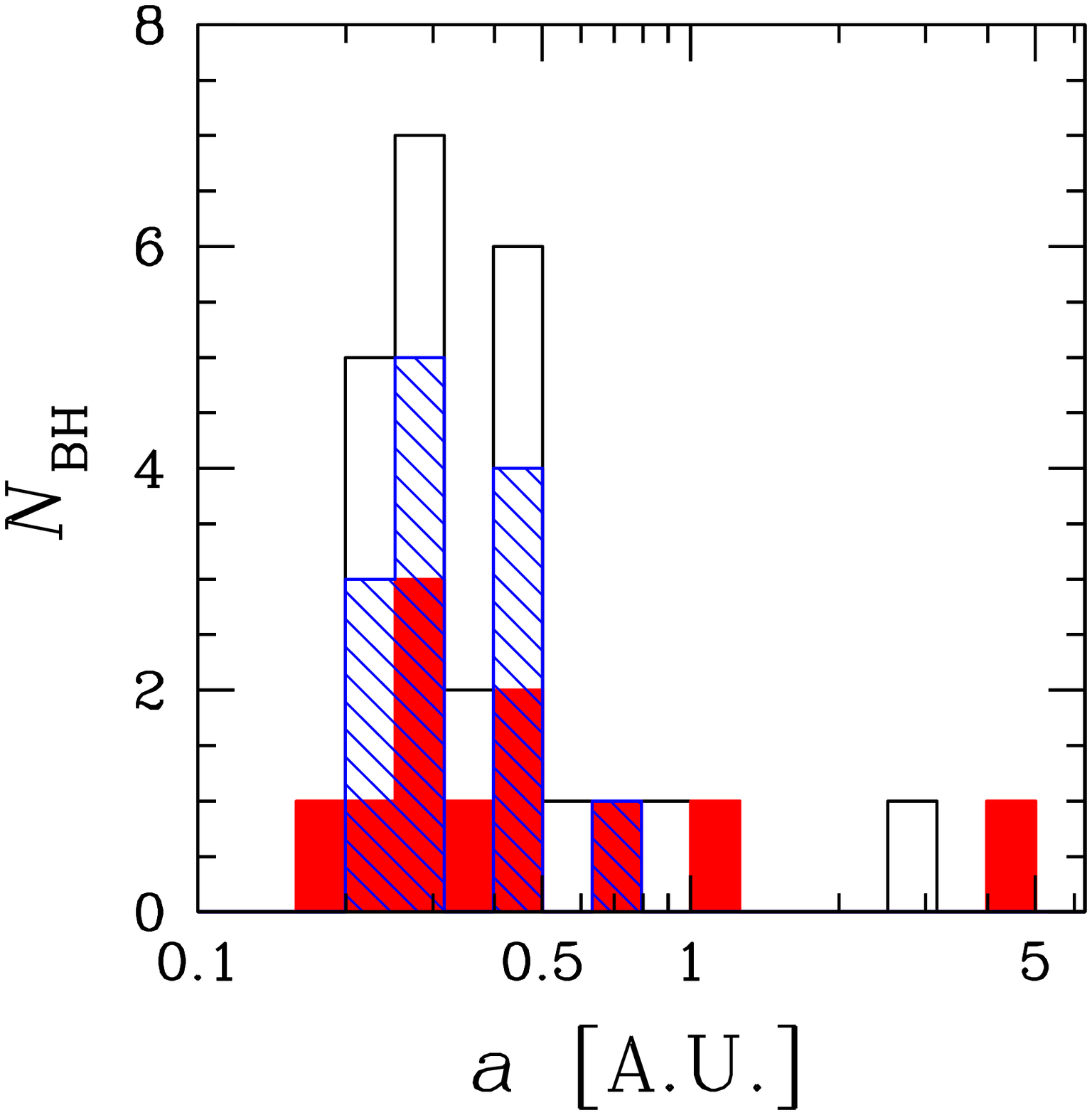,height=5cm} 
\epsfig{figure=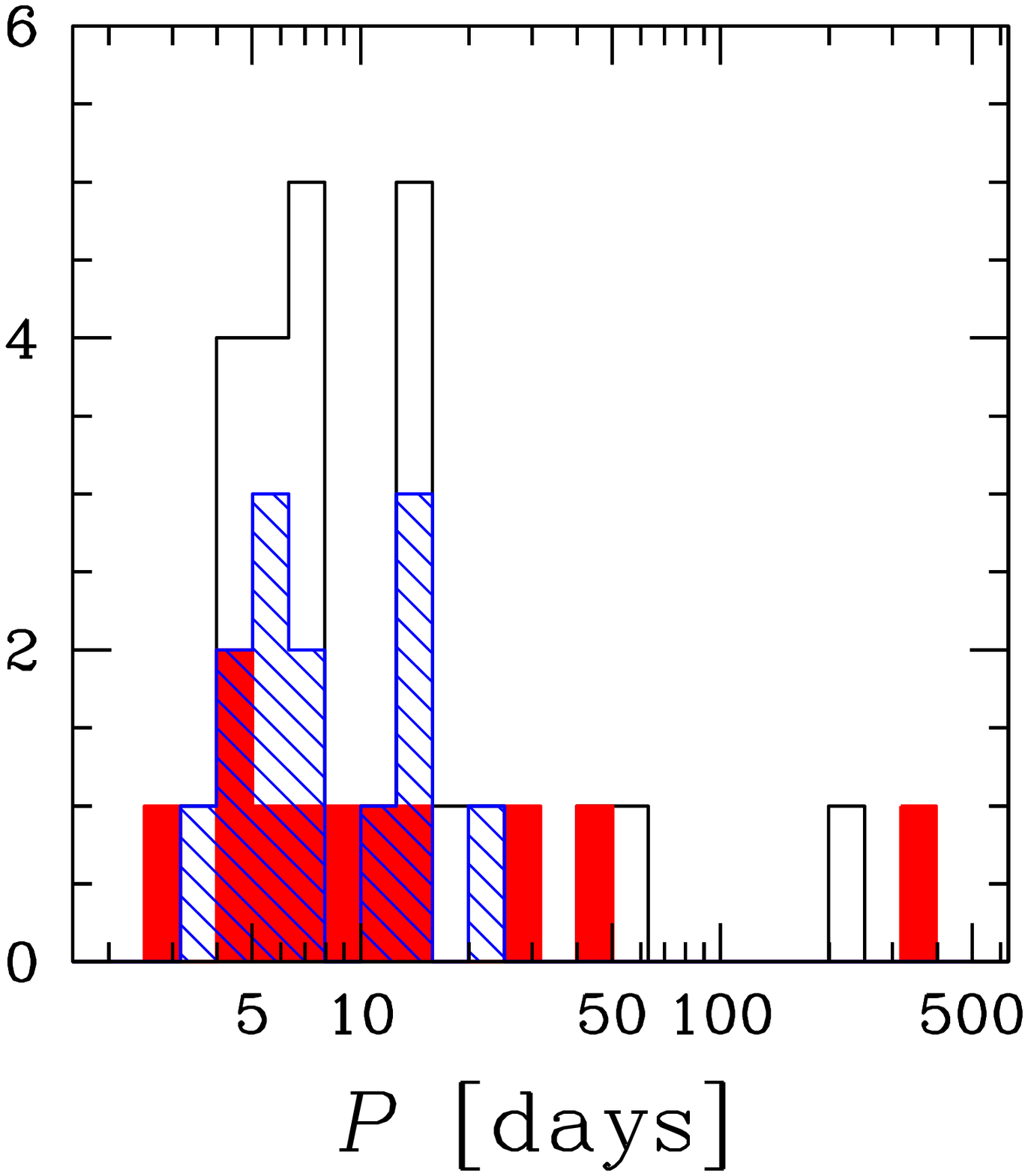,height=5cm} 
\epsfig{figure=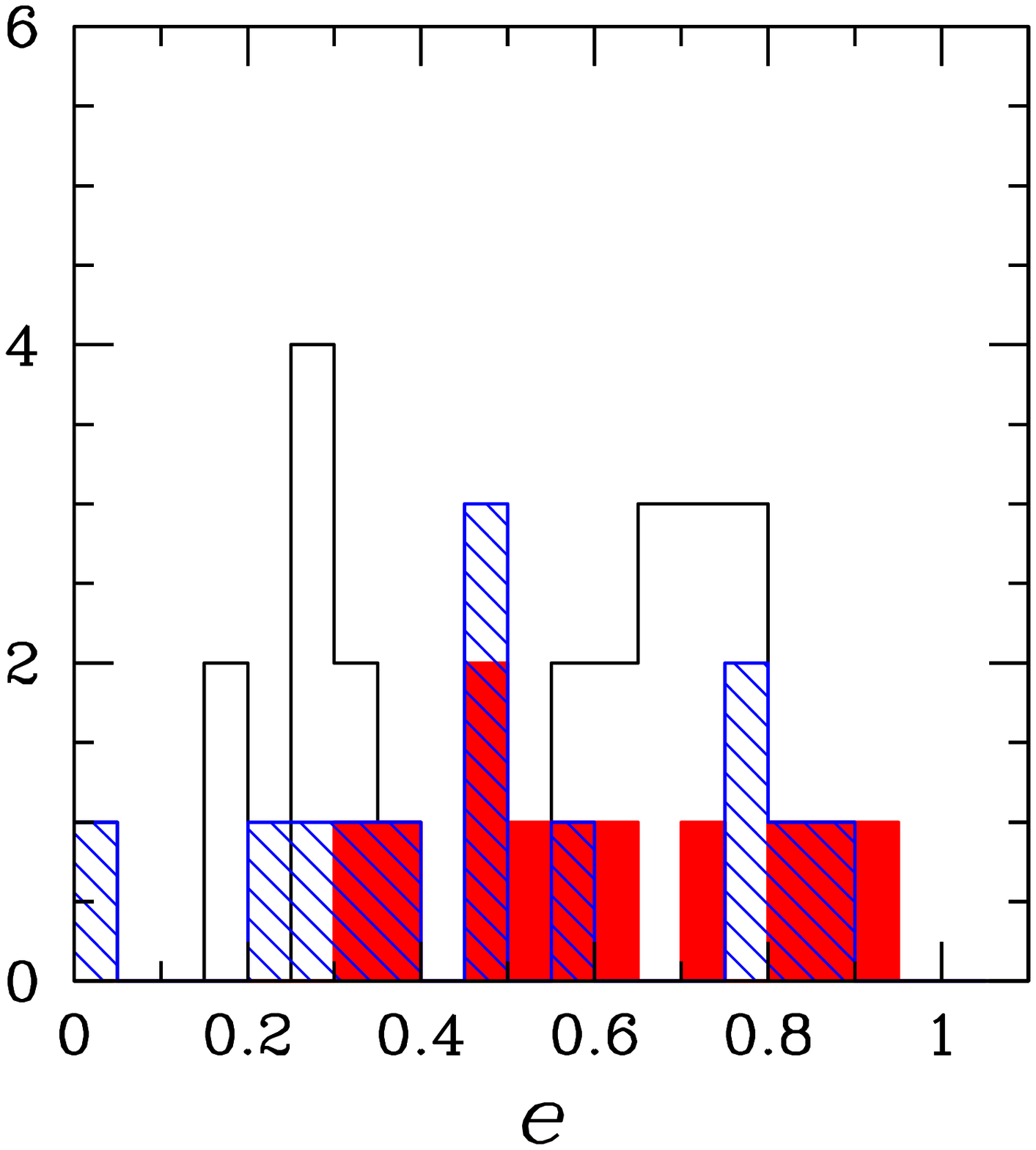,height=5cm} 
}}
\caption{\label{fig:fig5}
Semi-major axis $a$ (left-hand panel), period $P$ (central panel) and eccentricity $e$ (right-hand panel) from simulations of MSBH systems. In all the panels: the empty black histogram represents the initial conditions, the filled histogram (red on the web) shows the sub-sample of the ejected MSBHs at time $t=10$ Myr, whereas the hatched histogram (blue on the web) shows the sub-sample of the non-ejected MSBHs at time $t=10$ Myr. 
}
\end{figure*}
\begin{figure*}
\center{{
\epsfig{figure=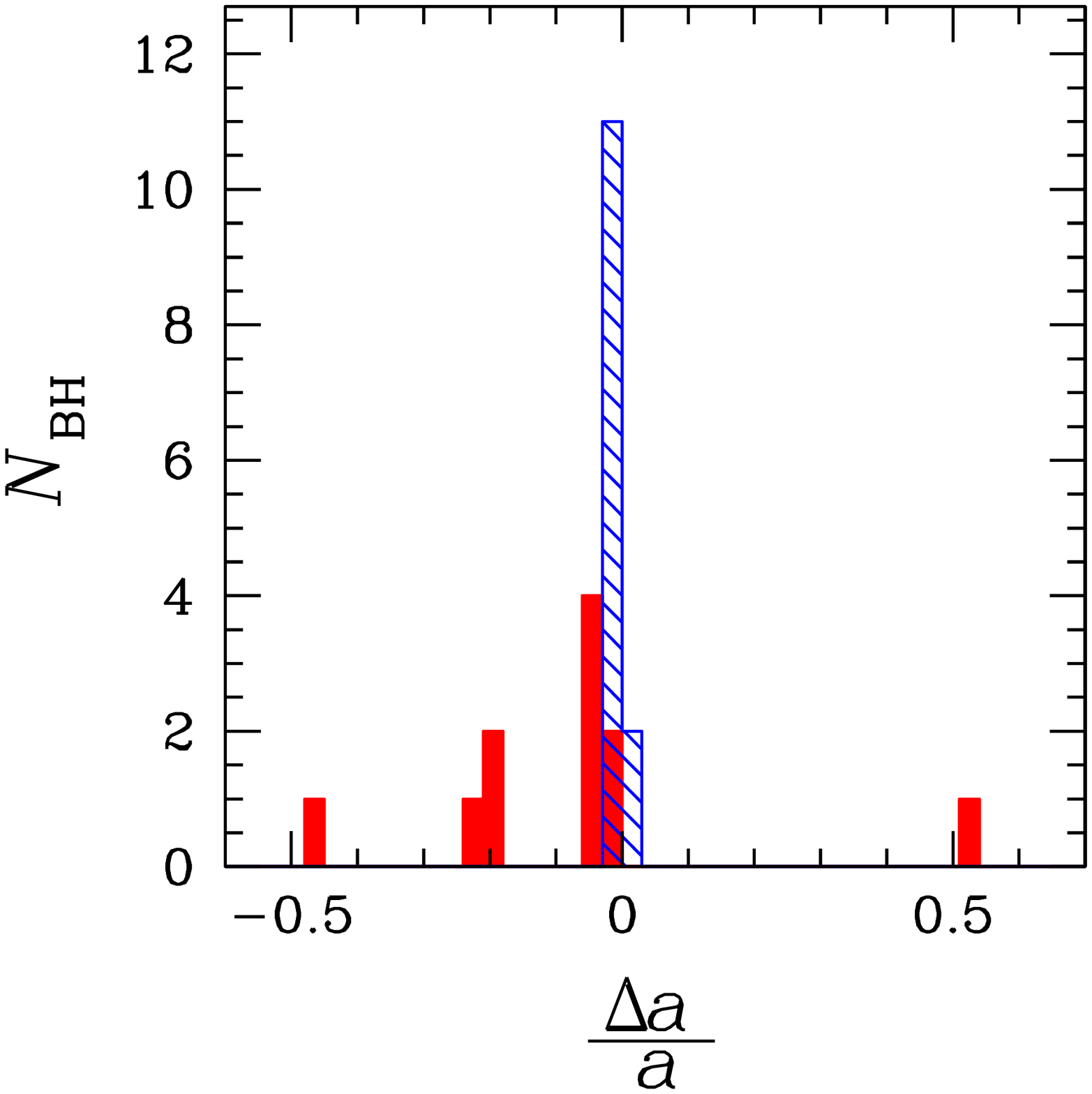,height=5cm} 
\epsfig{figure=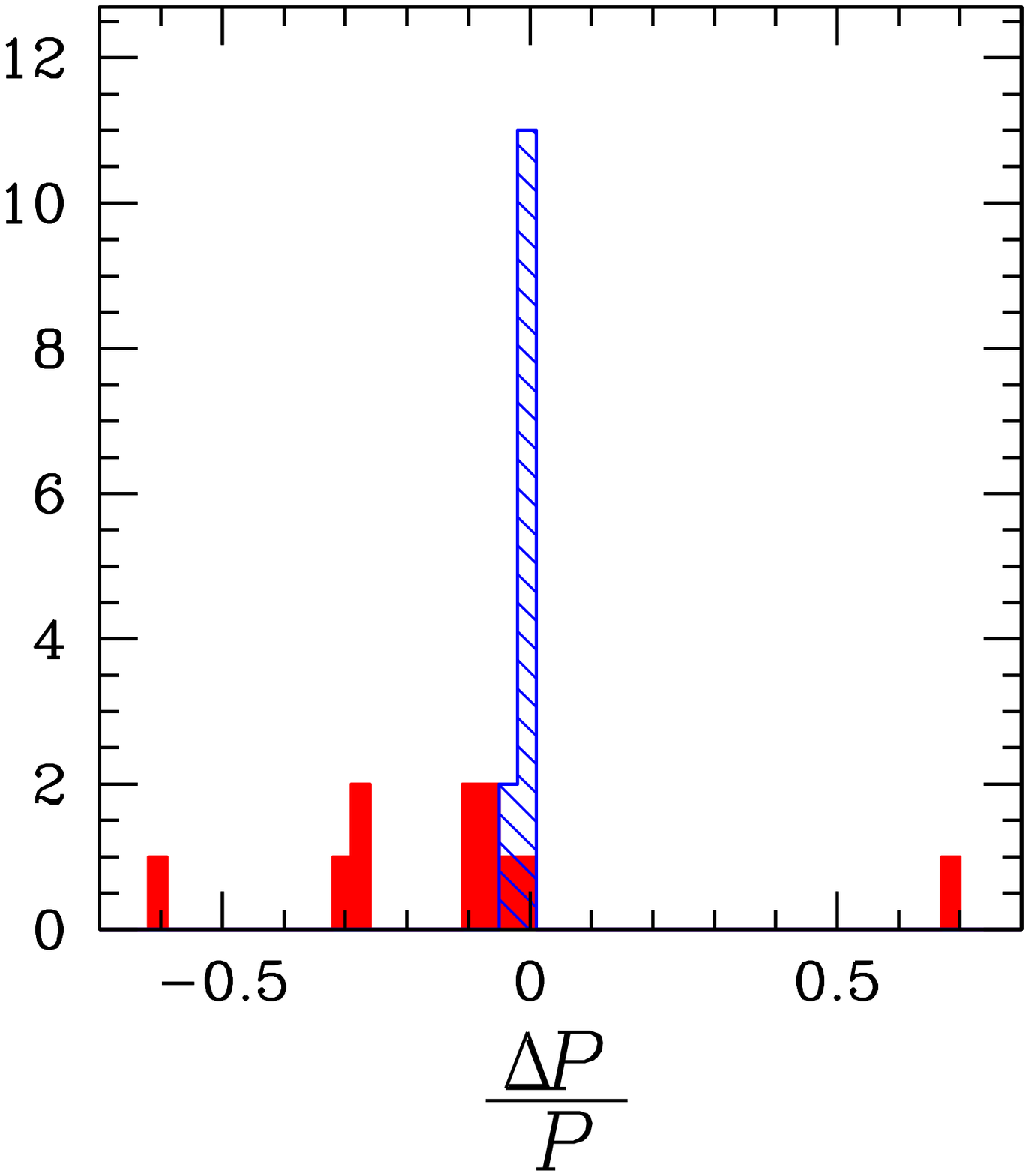,height=5cm} 
\epsfig{figure=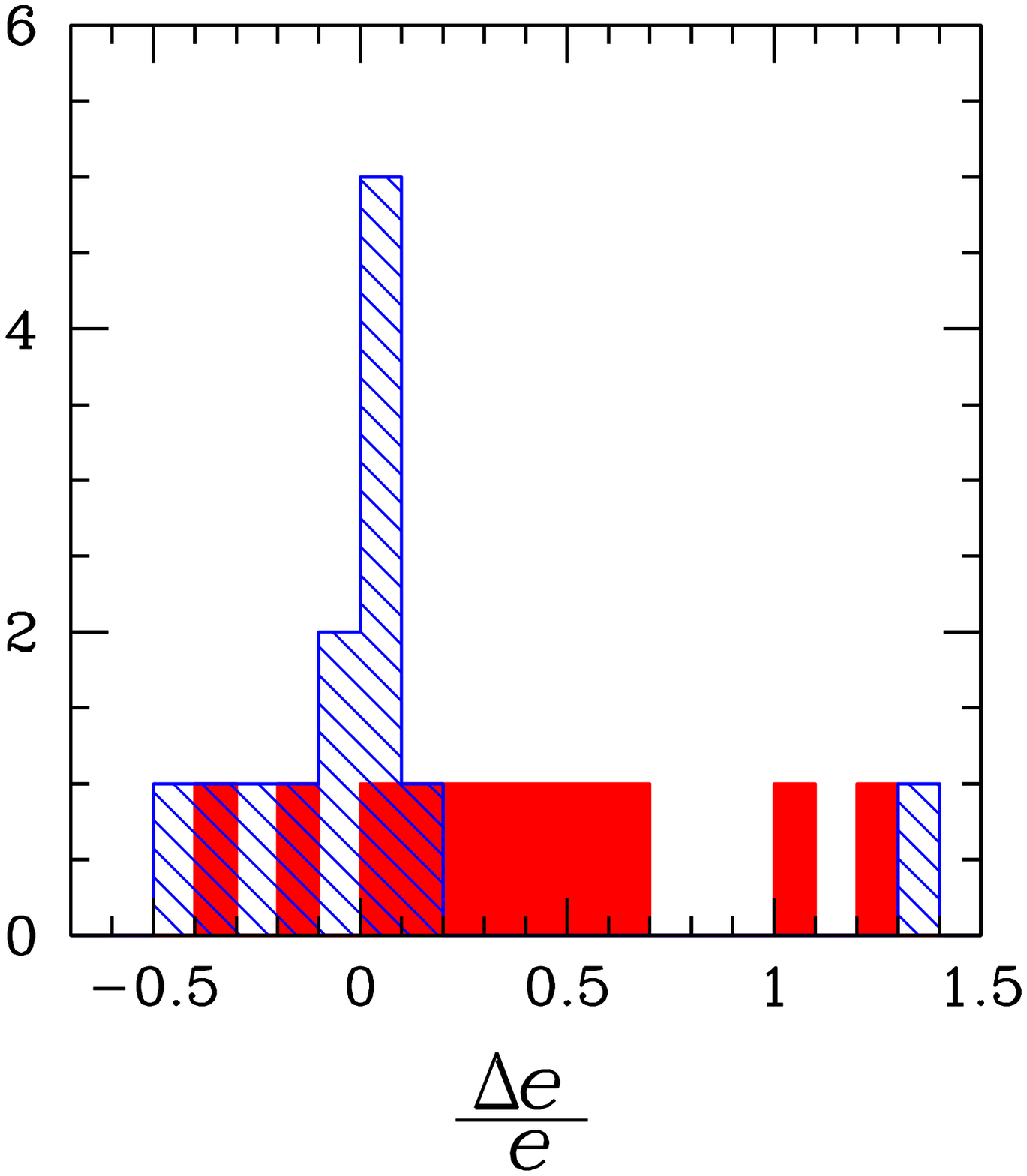,height=5cm} 
}}
\caption{\label{fig:fig6}
Fractional variation of the semi-major axis $a$ (left-hand panel), of the period $P$ (central panel) and of the eccentricity $e$ (right-hand panel) from simulations of MSBH systems. The filled histogram (red on the web) and the hatched histogram (blue on the web) show the sub-sample of the ejected MSBHs and that of the non-ejected MSBHs at $t=10$ Myr, respectively. 
}
\end{figure*}

\begin{figure}
\center{{
\epsfig{figure=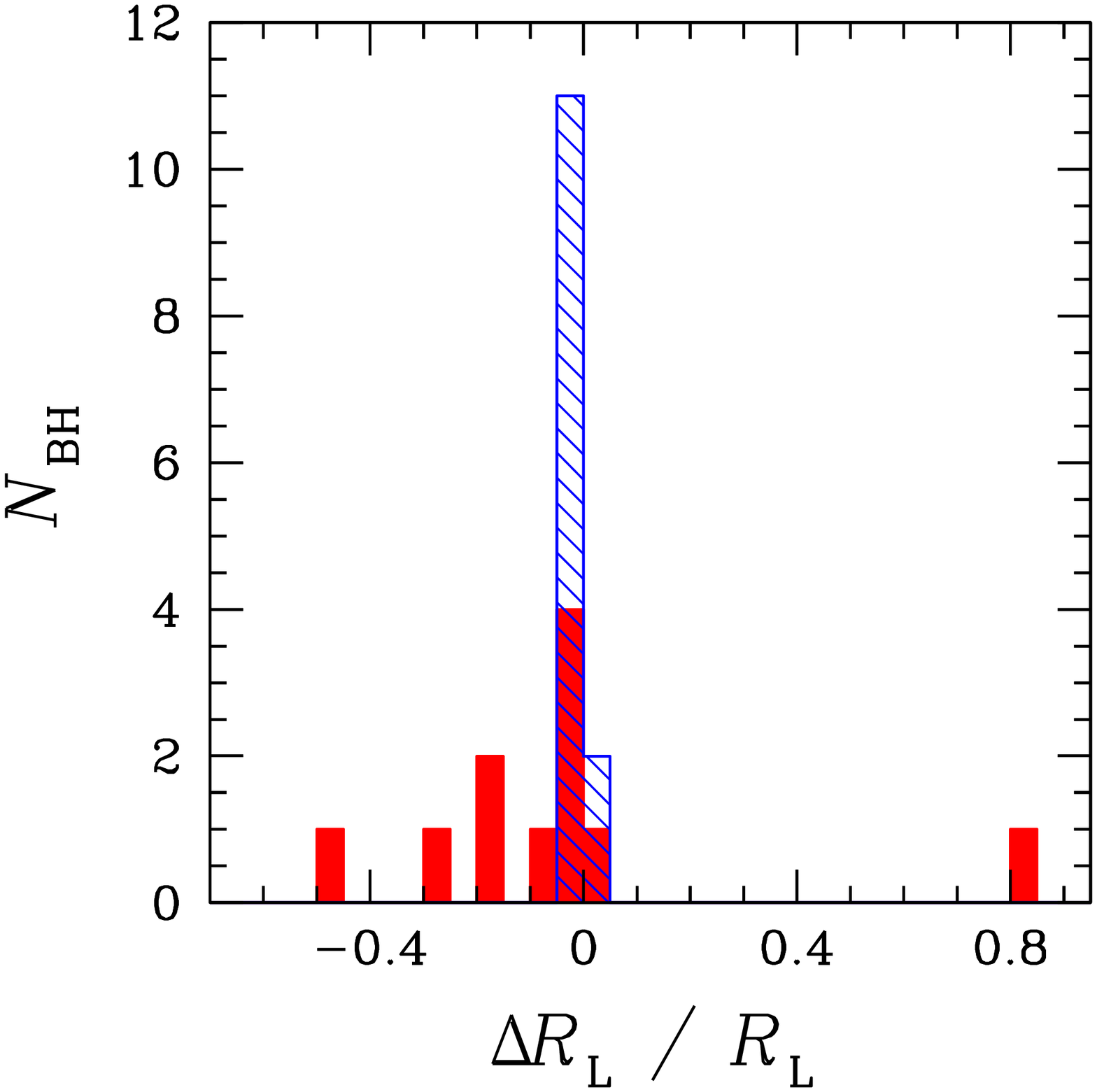,width=7cm} 
}}
\caption{\label{fig:fig7}
Distribution of the fractional changes of the Roche lobe ($\frac{\Delta{}R_{\rm L}}{R_{\rm L}}$). Hatched histogram (blue on the web): runs where the MSBH is still inside the parent cluster at $t=10$ Myr. Filled histogram (red on the web): runs where the MSBH is ejected before time $t=10$ Myr.}
\end{figure}

Fig.~\ref{fig:fig4} shows the distribution of times of MSBH ejection from the parent cluster. We calculate it with an accuracy of $\sim{}0.25$ Myr. Most of ejections occur between 2.5 and 6.5 Myr, when the massive companion of the MSBH is expected to be on the MS.

\section{Evolution of the orbital parameters}

Fig.~\ref{fig:fig5} shows the main orbital parameters of the simulated MSBH binaries, i.e., from left to right, semi-major axis $a$, period $P$ and eccentricity $e$. In all panels, the black empty histogram represents the initial conditions, whereas the ejected MSBHs (at $t=10$ Myr) and the MSBHs remaining in the cluster are represented by the filled histogram (red on the web) and by the hatched histogram (blue on the web), respectively.
The initial conditions illustrate what we mentioned in Section~2, i.e. the fact that we consider only hard binaries. A large fraction of the systems have periods typical of  RLO--HMXBs binaries ($P\lesssim{}10$ days). After 10 Myr, there is no big difference between ejected systems and binaries that remain inside the parent cluster. The only noticeable difference concerns the final distribution of eccentricities: after 10 Myr, the eccentricity of ejected binaries is, on average, larger than the eccentricity of systems that stay inside the cluster.  This agrees with the expectation that close encounters increase the eccentricity of a binary (see, e.g., Sigurdsson \&{} Phinney 1993; Rasio \&{} Heggie 1995; Heggie \&{} Rasio 1996; Colpi, Mapelli \&{} Possenti 2003).

In order to highlight the differences between ejected and non-ejected systems, it is better to look at the fractional change of the orbital parameters shown in Fig.~\ref{fig:fig6}. From left to right, the fractional change of semi-major axis ($\Delta{}a/a\equiv{}\left[a(t=10\,{}{\rm Myr})-a(t=0)\right]/a(t=0)$), period ($\Delta{}P/P\equiv{}\left[P(t=10\,{}{\rm Myr})-P(t=0)\right]/P(t=0)$) and eccentricity ($\Delta{}e/e\equiv{}\left[e(t=10\,{}{\rm Myr})-e(t=0)\right]/e(t=0)$) are shown. 
The main feature of non-ejected systems is that the fractional changes of semi-major axis and of period are extremely small ($|\frac{\Delta{}a}{a}|<0.02$ and $|\frac{\Delta{}P}{P}|\le{}0.025$). This indicates that non-ejected systems did not undergo a strong interaction. Instead, fractional changes of $a$ and $P$ for ejected systems are much  larger ($0.01\le{}|\frac{\Delta{}a}{a}|\le{}0.54$ and $0.01\le{}|\frac{\Delta{}P}{P}|\le{}0.7$), indicating a strong perturbation before the ejection. Furthermore, the semi-major axis decreases in most of the systems: $\frac{\Delta{}a}{a}<0$ for 91 and 85 per cent of ejected and non-ejected binaries, respectively. Again, this is consistent with the fact that the simulated binaries are hard and tend to become harder (Heggie 1975). It is interesting to note that only one ejected binary becomes wider and $\frac{\Delta{}a}{a}$ is quite large ($\sim{}0.5$). However, this system is quite peculiar, as it was by far the softest already in the initial conditions ($a(t=0)=2.8$ A.U.), it has been ejected late in time (at $t=9.25$ Myr), it is not far from the centre of the cluster (projected distance $r=6.8$ pc at $t=10$ Myr) and is one of the systems that underwent exchange of the companion.

The right-hand panel of Fig.~\ref{fig:fig6} shows the fractional change of eccentricity and strengthens the evidence that ejected binaries become more eccentric than non-ejected ones.

In general, three-body encounters change significantly the orbital parameters of MSBH binaries: this may be crucial for their evolution into RLO--HMXB systems. 
In fact, three-body encounters may have the same effects as natal kicks:
(i) they reduce the semi-major axis and can bring into RLO systems that, otherwise, would be too wide; (ii) they increase the eccentricity and may switch-on the RLO--HMXB phase, since a highly eccentric system is likely to be in RLO at periastron (and frictional forces rapidly circularize the orbits, leading to a stable RLO).


Thus, dedicated runs including stellar and binary evolution are required, to quantify the effect of gravitational encounters on mass transfer and to determine the onset of a RLO--HMXB phase in MSBH systems.
In this paper, we can only try to 
estimate the effects caused by changes
in the orbital parameters, by calculating the Roche lobe ($R_{\rm L}$). We adopt the most common approximation for the Roche lobe for circular orbits (see, e.g., Tauris \&{} van den Heuvel 2006):
\begin{equation}
\frac{R_{\rm L}}{a}=0.49\,{}\frac{q^{2/3}}{0.6\,{}q^{2/3}+\ln{}(1+q^{1/3})},
\end{equation} 
where $q=m_2/m_{\rm BH}$ (where $m_{\rm BH}$ is the mass of the MSBH and $m_2$ the mass of the companion).
Fig~\ref{fig:fig7} shows the fractional changes of the Roche lobe ($\Delta{}R_{\rm L}/R_{\rm L}\equiv{}\left[R_{\rm L}(t=10\,{}{\rm Myr})-R_{\rm L}(t=0)\right]/R_{\rm L}(t=0)$) for non-ejected (hatched histogram) and ejected MSBH binaries (filled histogram). We note that the Roche lobe of MSBHs remaining inside the parent cluster is essentially unchanged, whereas the Roche lobe of ejected MSBH binaries may significantly decrease ($|\frac{\Delta{}R_{\rm L}}{R_{\rm L}}|\le{}0.5$). This is a direct consequence of the shrinking of the binary after a three-body encounter (see the left-hand panel of Fig.~\ref{fig:fig6}), combined with the possible change of $m_2/m_{\rm BH}$ through dynamical exchanges. Dynamical exchanges may be crucial for entering the RLO regime, as they generally lead to binary systems with more massive companion stars. 
This suggests that the orbital shrinking may switch-on RLO in ejected binaries, although our simulations cannot be conclusive, because of the small statistics and  the absence of stellar evolution calculations\footnote{Because of the absence of stellar evolution in our simulations, we cannot estimate the radius of the companion star $r_\ast{}$. Therefore, we do not know the value of the ratio $R_{\rm L}/r_\ast{}$ and we cannot quantify whether a simulated system undergoes RLO. However, for reasonable assumptions about the stellar radius-mass relation (see, e.g., Prialnik 2000), $R_{\rm L}/r_\ast{}\approx{}0.8-20$ in the initial conditions. $R_{\rm L}/r_\ast{}$ remains almost constant for the retained binaries. The number of ejected binaries with $R_{\rm L}/r_\ast{}\lesssim{}1$ is four at $t=10$ Myr (only two of them had $R_{\rm L}/r_\ast{}\lesssim{}1$ at $t=0$).}.

\section{Comparison with IMBHs}
For comparison with the MSBHs, we run 20  simulations of clusters having the same properties as before, but hosting a 300 M$_\odot{}$ IMBH. The adopted mass is quite high for some scenarios of IMBH formation. For example, Portegies Zwart \&{} McMillan (2002) indicate that the mass of an IMBH formed via runaway collapse should be about 0.1 per cent of the mass of the parent cluster. Actually, the prediction by Portegies Zwart \&{} McMillan (2002) should likely be considered an upper limit, as it does not include recipes for mass loss during the collapse (e.g., Gaburov, Lombardi \&{} Portegies Zwart 2010).  Thus, our assumption for the mass of the IMBH is somewhat unrealistic. On the other hand, it would be highly expensive to simulate a statistically significant sample of  $3\times{}10^5$ M$_\odot{}$ clusters with the required accuracy. Furthermore, the density in the core of the simulated clusters ($10^4-10^5$ stars pc$^{-3}$) is not far from the density of more massive clusters and the dynamical evolution of the IMBH is expected to be quite similar in the two environments.

\begin{figure}
\center{{
\epsfig{figure=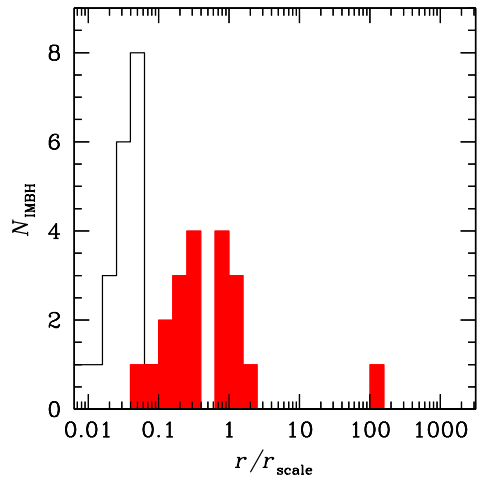,height=7cm} 
}}
\caption{\label{fig:fig8}
Empty black histogram: two-dimensional radial projected distribution of IMBHs at time $t=0$. Filled histogram (red on the web): two-dimensional  radial projected distribution of IMBHs at time $t=10$ Myr. $y-$axis: number of simulated IMBHs ($N_{\rm IMBH}$); $x-$axis: two-dimensional distance of the IMBH from the centre of the parent cluster $r$, normalized to the scale radius $r_{\rm scale}$. For all the simulated clusters  $r_{\rm scale}\sim{}1$ pc.
}
\end{figure}

Fig.~\ref{fig:fig8} shows the two-dimensional projected distance of IMBHs from the centre of the parent cluster at $t=0$ (empty histogram) and $t=10$ Myr (filled histogram).  The bottom panel of Fig.~\ref{fig:fig9} shows the projected distance of IMBHs at $t=10$ Myr (hatched histogram), in comparison with the projected distance of MSBHs at $t=10$ Myr (filled histogram). First, we note that only one IMBH (5 per cent of the sample) has been ejected from the cluster after $t=10$ Myr. This is not surprising, as we expect recoil velocities for IMBHs to be at least a factor of 3 lower than those for MSBHs. We note that the distance of IMBHs from the centre of the parent cluster at $t=10$ Myr is larger than in the initial conditions. This was true also for MSBHs (see Fig.~\ref{fig:fig2}) and is likely due to the fact that three-body encounters push into the outskirts of the cluster even those BHs that were not completely ejected (as predicted, e.g., by Kulkarni et al. 1993). We stress that this migration of BH binaries to the outskirts does not affect the global distribution of stars in the cluster, as the overall profile does not change significantly (see Fig.~\ref{fig:fig1}). It must be noted that MSBHs and IMBHs that are pushed into the periphery of the parent cluster might be more easily stripped from it, by external tidal fields (e.g. the one of the host galaxy). On the other hand, the escape of IMBHs and MSBHs from the parent cluster, because of tidal fields, implies (i) longer timescales than direct three-body ejection (likely longer than the lifetime of a massive companion star); (ii) the lack of recoil velocity.
Thus, the simulations confirm that IMBHs can hardly be ejected from the parent cluster by three-body encounters. 


\section{Comparison with data}
\begin{figure}
\center{{
\epsfig{figure=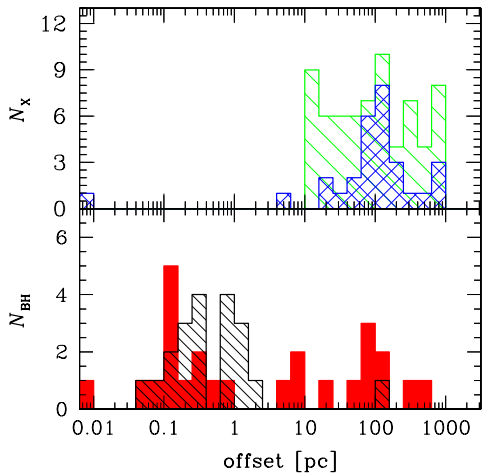,height=7cm} 
}}
\caption{\label{fig:fig9}
Bottom panel: two-dimensional radial projected distribution of IMBHs (hatched histogram) and of MSBHs (filled histogram, red on the web) at time $t=10$ Myr. 
 Bottom $y-$axis: number of simulated BHs ($N_{\rm BH}$); $x-$axis: two-dimensional distance of the BH from the centre of the parent cluster (offset) in pc. 
Top panel: the cross-hatched histogram (blue on the web) shows the radial projected distribution of observed ULXs from the closest star-forming region (from the sample in table~9 of Berghea 2009); the hatched histogram (green on the web) shows the radial projected distance of observed X-ray sources (with X-ray luminosity $L_X\ge{}5\times{}10^{35}$ erg s$^{-1}$) from the closest star cluster (considering M~82, NGC~1569, NGC~5253, from figure 3 of Kaaret et al. 2004b). Top $y-$axis: the number of observed X-ray sources ($N_{\rm X}$); $x-$axis: projected distance of the X-ray source from the closest star cluster and/or star forming region, in pc.
}
\end{figure}

Fig.~\ref{fig:fig2} compares the simulations of MSBHs with
two different observational samples (see also Fig.~\ref{fig:fig9}, for a comparison of the two observational samples with the simulations of both MSBHs and IMBHs). The hatched histogram (green on the web) shows the offset of bright X-ray sources (with X-ray luminosity $L_{\rm X}\ge{}5\times{}10^{35}$ erg s$^{-1}$) from the closest star cluster in a sample of three nearby starburst galaxies: M82, NGC~1569 and NGC~5253 (from fig.~3 of Kaaret et al. 2004b). The observed displacements reported by Kaaret et al. (2004b) range from $\sim{}10$ pc to $\sim{}1000$ pc, in good agreement with the projected positions of ejected MSBHs in our simulations (filled histogram). We note that displacements smaller than 10 pc have been set to 10 pc by Kaaret et al. (2004b), because their spatial uncertainty is about 10 pc. Therefore, nothing can be said about X-ray sources inside the parent cluster, on the basis of Kaaret et al. (2004b).

The cross-hatched histogram (blue on the web) shows the offsets of a sample of ULXs from the closest region of star formation (data come from table~9 of Berghea 2009). Although the offsets reported by Berghea (2009) refer generically to star forming regions (which cannot be distinguished from YSCs), the superposition between the position of simulated ejected MSBHs and the data is remarkable. Both peak at $\sim{}100$ pc and range from $\sim{}4$ pc up to $\sim{}1000$ pc. Only the position of a single ULX (IC~342 X-3) in Berghea (2009) is found to be completely coincident with the centre of a star forming region. This might be at odds with the fact that more than half of the simulated MSBHs at $t=10$ Myr are still inside the parent cluster, but there are various possible explanations for this difference.

First, data by Berghea (2009) represent the offset between the ULX and the centre of the closest star forming region, identified on the basis of the colour; but each star forming region may host more than one YSC and/or OB association. Thus, the position of the ULX might be coincident with that of a YSC enclosed in the star forming region, but not identified by  Berghea (2009).  Berghea (2009) highlights that about half of the ULXs in his sample are inside the star forming region, although displaced with respect to its centre. Furthermore, table~8 of Berghea (2009) indicates that, out of a sample of 31 ULXs, 6 have no detected counterpart (2 of them do not even have a star forming region within 1 kpc), 12 have single-star candidates as counterparts and 13 have counterparts that are brighter than a single O5V star. Among the latter, 5 ULXs are associated with ultra-violet (UV) emission consistent with $\ge{}14$ O5V stars, which possibly means YSCs similar to those we simulated (although the UV excess may be due in part to contamination from the accretion disc).
In particular, Berghea (2009) highlights that Holmberg IX X-1 is located at the edge of a stellar cluster with estimated mass of $10^3$ M$_\odot{}$ (Gris\'e, Pakull \& Motch 2006), NGC~1313 X-2 is located in the periphery of a relatively young cluster (Zampieri et al. 2004), IC~342 X-3 is inside the nuclear cluster (B\"oker, van der Marel \&{} Vacca 1999) and NGC~4449 X-7 is associated with a compact young cluster (Gelatt, Hunter \&{} Gallagher 2001; Soria et al. 2005). Therefore, the data by Berghea (2009) do not necessarily exclude the association of  a fraction of ULXs with YSCs.

On the other hand, there may be intrinsic reasons why HMXBs from MSBHs are suppressed inside star clusters. 
 For example, a MSBH binary that remains inside the cluster continuously undergoes three-body interactions (see Table~2) and cannot evolve unperturbed. This might strongly affect the mass transfer (see, e.g., Mapelli et al. 2004, 2006 for the effect of dynamical perturbations onto blue straggler formation from mass-transfer binaries). These effects might be very important and need to be studied with dedicated binary-evolution simulations.

\section{Conclusions}
In this paper, we studied the dynamical evolution of hard binaries hosting MSBHs with massive companions. We showed that a large fraction ($\sim{}44$ per cent) of MSBHs is ejected from the cluster on a short timescale ($<10$ Myr), because of close interactions. This result is very different from what the simulations indicate for IMBHs: only a small fraction of IMBHs ($\sim{}5$ per cent) is completely ejected from the parent cluster.

All the ejected MSBHs retain their companion, even after the ejection. Furthermore, at $t\sim{}10$ Myr, the offset of the ejected MSBHs with respect to the centre of the parent cluster is consistent with the observations of Kaaret et al. (2004b) for HMXBs and with those of Berghea (2009) for ULXs. These results also suggest that dynamical recoil for MSBHs has similar effects to natal kick for stellar BHs. Thus, assuming that MSBHs born from direct collapse have no (or negligible) natal kick, the dynamical recoil is the only efficient mechanism to eject MSBH binaries from their parent cluster.

 In some cases (especially for IMBHs), the recoil due to three-body encounters is not sufficient to promptly eject the BH, but the BH is pushed into the outskirts of the cluster. If this occurs, the BH might escape at a later stage, because of the influence of the galactic tidal field or cluster evaporation/disruption.

Also, three-body interactions might affect the orbital parameters of the MSBH binaries. The semi-major axis and, consequently, the period change sensibly after the ejection. On average, the binary hardens, and experiences a fractional decrement of the semi-major axis of $0.01-0.5$.
Ejected MSBH binaries tend to become more eccentric. 
These effects on the orbital parameters of MSBH binaries might have important consequences on the occurrence of mass transfer. In fact, a decrease of the semi-major axis might bring into RLO systems that were, otherwise, too far apart. 

Recent studies (Mapelli et al. 2009; Zampieri \&{} Roberts 2009; Mapelli et al. 2010a,b) propose that MSBH binaries might power a  fraction of bright HMXBs and ULXs. On the other hand, L10 indicate that MSBH binaries can hardly become RLO--HMXBs, because of the lack of natal kick in their formation pathway. 
In this paper, we suggest that dynamical interactions may have similar effects as the natal kick. This argument is strengthened by the agreement between the radial distribution of simulated MSBHs and the offset observed in many ULXs and/or HMXBs with respect to nearby star clusters and/or star forming regions. We also suggest that dynamical interactions can change the orbital parameters of MSBH binaries, favouring mass transfer. Thus, it will be crucial to assess whether dynamical interactions may effectively lead MSBH systems to pass through RLO--HMXB phase.
For this reason, dedicated binary evolution calculations, coupled with dynamical simulations, are definitely needed.





 Furthermore, the aforementioned papers (Mapelli et al. 2009; Zampieri 
\&{} Roberts 2009; L10;  Mapelli et al. 2010a,b) consider only the formation 
of massive BHs as remnants of  metal-poor stars. However, there might be 
different pathways for the formation of massive BHs, even independent of the 
metallicity. For example, in dense star clusters, 
massive BHs  are expected to form through (single or multiple) mergers of hard binaries, as a consequence of stellar evolution (e.g., Belczynski et al. 2004), 3-body encounters (e.g., Miller \& Hamilton 2002) and/or (in case of binaries of compact objects) gravitational wave emission. In addition, the runaway collapse process (e.g., Portegies Zwart \&{} McMillan 2002; G\"urkan, Freitag \&{} Rasio 2004; Freitag, G\"urkan \&{} Rasio 2006), which is expected to lead to the formation of IMBHs, can be interrupted when the BH mass is still $\le{}100$ M$_\odot{}$, because of the dynamical ejection of the BH seed (e.g., Portegies Zwart \&{} McMillan 2000; Miller \&{} Colbert 2004) or of strong mass losses in the merger phase (e.g., Gaburov, Lombardi \&{} Portegies Zwart 2010).
Such massive BHs born from an interrupted runaway collapse or from binary mergers need further investigation.

\section*{Acknowledgments}
We thank the anonymous referee for the useful comments and C.~Berghea for allowing us to use the results of his PhD Thesis. We thank A.~Bressan and P.~Marigo for useful discussions. We made use of the public software package Starlab (version 4.4.4). We acknowledge all the developers of Starlab, and especially its primary authors: Piet Hut, Steve McMillan, Jun Makino, and Simon Portegies Zwart. 


\end{document}